\documentstyle[12pt,aasms4]{article}

\begin{document}

\title{``Baade's red sheet'' resolved into stars
with HST\footnotemark[1] \ in the Blue Compact Dwarf Galaxy VII~Zw~403 }

\author{R.E. Schulte-Ladbeck, Mary M. Crone\footnotemark[2]}
\affil{University of Pittsburgh, Pittsburgh, PA 15260}
\authoremail{rsl@binar.phyast.pitt.edu}
\author{U. Hopp}
\affil{Universit\"{a}tssternwarte M\"{u}nchen}

\footnotetext[1]{Support
for this work was provided by NASA through grant
number AR-06404.01-95A from the Space Telescope Science Institute,

which is operated by the Association of Universities for Research in
Astronomy,
Incorporated, under NASA contract NAS5-26555.}

\footnotetext[2]{Current address:  Skidmore College, Saratoga Springs,
NY 12866}

\begin{abstract}

HST WFPC2 observations of the nearby Blue Compact Dwarf Galaxy
VII~Zw~403 (= UGC~6456) resolve single stars down to
M$_I$$\approx$-2.5, deep enough to identify red giants. 
This  population has a more
uniform spatial distribution than the young main-sequence stars and
supergiants, forming the structure known as ``Baade's red sheet".  We
conclude that VII~Zw~403 is not a primeval galaxy.

\end{abstract}


\section{Introduction}

In 1972, Searle \& Sargent announced that two prototypical Blue
Compact Dwarf Galaxies (BCDs) were ``the first metal-poor systems of
Population I to be discovered''.  They posed the question of whether
BCDs are young galaxies in the sense that most of their star formation
happened recently, or old galaxies in which star
formation occurs in intense bursts separated by long quiescent
intervals.  

An old galaxy must show evidence for the presence of an old stellar
population. Since BCDs are generally too small and too distant to
resolve into stars from the ground, studies of their stellar
populations have been based on analyses of global galaxy colors and
spectra. In his 1991 review, Thuan discussed several observations
which, taken together, made him favor the old-galaxy hypothesis.
Based on spectral synthesis modeling of IUE spectra, Fanelli et al.
(1988) found that the star-formation history of BCDs is best
characterized by multiple, discrete star-forming episodes rather than
by a single burst or a continuous star-formation rate. Thuan (1983)
obtained near-infrared photometry of BCDs and argued he had found an
old population of red giants.  Unfortunately, as Thuan recognized, it
is difficult to discriminate between a population of young red
supergiants and one of old red giants using only the total infrared
colors of a galaxy. Campbell \& Terlevich (1984) obtained CO indices
and asserted that the population detected in the infrared is primarily
composed of supergiants from the current starburst.  Loose \&
Thuan (1986), Kunth et al. (1986), Kunth et al. (1988) and Papaderos
et al. (1996) studied the optical morphology of BCDs.  They found that
many BCDs show more extended and regular morphologies in red images
than in blue images, with colors becoming redder with increasing
distance from the starburst centers. These observations have been
interpreted to indicate the presence of ``Baade's red sheet'', the
signature of an old stellar population, in some BCDs. However, there
are BCDs which do not appear to exhibit these extended red halos.

Three BCDs of been observed with HST/WFPC2, two of
which are extremely metal poor and thus good candidates for primeval
galaxies.  But because these BCDs are at distances greater than
10~Mpc, even the extremely high spatial resolution of HST could not 
resolve any faint old stars in these galaxies.  Mrk~966,
studied by Thuan et al. (1996), has an oxygen abundance of 1/10 of
solar. This BCD is somewhat resolved: Thuan et al. find 
$\sim$~40 point-like sources around and projected onto an unresolved disk,
which they identify with an old globular cluster system.  SBS~0335-052
has an oxygen abundance of 1/40 of solar. Thuan et al.  (1997) find
young super-star clusters of this galaxy are resolved, but its disk 
is not. The irregular morphology and blue color of the disk
lead Thuan et al. to suggest that this is probably a young galaxy. In
I~Zw~18, the most metal-poor galaxy on record with an oxygen abundance
estimated to be between 1/40 to 1/60 of solar, Hunter \& Thronson
(1995) successfully resolve individual massive stars with WFPC2.
However, any evolved descendants of lower-mass stars are below the
detection limit; once again, information about the galaxy's youth is
limited to the integrated colors of the disk.

VII~Zw~403 has recently attracted attention with the ROSAT/PSPC
discovery of an extended hot gas outflow powered by the present
starburst (Papaderos et al. 1994). The oxygen abundance of VII~Zw~403
is about 1/15 of solar; its HI mass is 2x10$^8$~M$_\odot$ (Tully et
al. 1981).  Prior to the observations presented in this paper, several
pieces of evidence suggested an old population might be present in
VII~Zw~403.  It belongs to the morphological type iE in the
classification of Loose \& Thuan, which designates irregular
star-forming centers surrounded by an elliptical
low-surface-brightness envelope. The red color of the unresolved disk
is interpreted to indicate an old stellar population (Schulte-Ladbeck
\& Hopp 1997). Carozzi et al. (1974) presented an optical spectrum of
VII~Zw~403 which displays a continuum break in the area of the Ca~H
and K and the G-band absorption features. VII~Zw~403 resolves into
single stars and/or star clusters from the ground (e.g., Fisher \&
Tully 1979, Tully et al. 1981, Karachentsev et al. 1994, Hopp \&
Schulte-Ladbeck 1995), but only a few of the most luminous objects can
be discerned.

We here present WFPC2 observations of the most nearby of the
well-studied BCDs (Kunth \& S\`{e}vre 1986), VII~Zw~403~=~UGC~6456.
We demonstrate that by choosing a BCD that is nearby, we can use HST
to resolve ``Baade's sheet'' into single stars. We thereby show
conclusively that VII~Zw~403 is not a young galaxy.

\section{Observations \& Data Reduction}

The HST WFPC2 observations of VII~Zw~403 are summarized in Table 1.
We re-ran the pipeline calibration with improved
calibration files, 
removed cosmic rays with CRREJ, corrected for CTE using a
4\% ramp, and corrected for geometric distortion using the image
available from STScI.

Plate 1 displays the three-color (F814W, F555W, and F336W) image of
the PC chip, illustrating the morphology of the central region of
VII~Zw~403.  The blue stars are clumped together, and are often
associated with nebulosity, whereas the faint red stars form a nearly
uniform sheet across the image.  There are several thousand point
sources in the PC alone (Table 1)!  
In the interest of processing all stars
identically and limiting contamination by foreground and background
sources, we include only stars in the PC in our analysis. 

We performed photometry on H$\alpha$-subtracted images.  Subtracting
H$\alpha$ had three effects. First, a few stars are line-emission
sources; these decreased in brightness by as much as a magnitude.
Second, many of the fainter stars ($m_{F555W} < 26$) that were
detected at the 3$\sigma$ level in the images without H$\alpha$
subtraction are no longer detected at this level, primarily 
due to increased noise. This changes the appearance of the
``red tangle'' (see below).  Third, the fainter stars which were still
detected in F555W became slightly brighter (by less than 0.1 mag for
$m_{F555W}< 26$), probably thanks to better sky subtraction.

We used DAOPHOT/ALLSTAR to do psf fitting for the crowded field.  We
chose relatively isolated stars in the image to determine the psf,
after finding that this produced a better fit than using TinyTim or
independent observations.  We calibrated the photometry with the May
1997 updated SYNPHOT tables, and used the Cool \& King (1995) method
to maximize the performance of DAOPHOT.  ALLSTAR residuals for the
F555W and F814W filters remain below 0.1~mag for magnitudes less than 25.5. 
For F336W the residuals exceed 0.1~mag
at about 24.0~mag.

We checked the completeness of our photometry 
for each  H$\alpha$-corrected image
by adding a distribution of false stars consistent with the magnitude
distribution of the real stars.  We ran 100 simulations for each
image, adding 10\% of the number of real stars each time, in order to
maintain the approximate level of crowding in the original image.  The
percentage of recovered stars indicates that there is no significant
incompleteness brighter than m$_{F555W}$~=~26, m$_{F814W}$~=~25, 
and m$_{F336W}$~=~25. 

Foreground galactic reddening for VII Zw403 is E(B-V)~=~0.025
(Burstein \& Heiles); we correct for the corresponding extinction
using the tables provided in Holtzman et al. (1995).  Tully et
al. (1981) find large, anomalous internal extinction near the large
southern HII-region, which is included in the PC image (see Plate 1).
However, in our color-color diagram (cd. Fig. 1), we see no
significant reddening for most stars.  
Therefore, we applied no
correction beyond the one for foreground extinction.  
It is quite possible that, 
as in other metal-poor dwarf galaxies (Calzetti et al. 1997), 
reddening is much larger for the ionized gas
than for the diffuse stellar population. 
We transformed to U, V, and I magnitudes using the color terms
in Holtzman et al. We note that there are significant uncertainties
in the transformation to U, which limit the accuracy of our extinction 
estimate, but do not change the shape of the [(V-I), I] 
color-magnitude diagram (CMD).  


Figure 1 shows color-magnitude diagrams for stars in the PC. We
include all objects detected to at least 3$\sigma$ which were
successfully fit to the psf.  Given the very small size of the
PC and the latitude $l=37.3$ of VII~Zw~403, the models of 
Ratnatunga \& Bahcall (1985) predict negligible foreground contamination. 

\section{Discussion}

\subsection{The Tip-of-the-Red-Giant-Branch Distance to VII~Zw~403}

Lee et al. (1993) showed that the tip of the red giant branch (TRGB)
consistently occurs at an absolute I Magnitude M$_I$$\approx$-4, with
only a slight dependence on metallicity.  We use this method to
determine the best distance estimate to date for VII~Zw~403.

Figure 2 shows our I-band luminosity function for the red stars only;
to bring out the red giant branch, we include only stars redward of
(V-I)$_o$=0.7.  The TRGB is so obvious as a sudden rise at I$_o$=24.3 that
we did not need to use an edge-finding algorithm.  (Other sources of
error are much larger!)  
Following the method of Lee et al., we  need a
value of [Fe/H] to determine M$_{bol, TRGB}$. It is difficult to
determine [Fe/H] from the color of the ``red tangle'', however, any
reasonable value for [Fe/H] does not change M$_{bol, TRGB}$ by a large
amount.  Using (V-I)$_o$$\approx$1.5 and a metallicity [Fe/H] of -1.2,
we find an absolute magnitude for the TRGB in VII~Zw~403 of
M$_I$$_o$=4.10.  
Our error budget for the distance modulus is summarized in Table 2.
The largest source of error is that from the RR Lyrae distance
calibration of the TRGB, which we take to be 0.15 mag, following Sakai
et al. (1997).
The distance modulus is
28.4$\pm$0.09$\pm$0.16, yielding a distance of
$4.8^{+0.4}_{-0.5}$~Mpc.

\subsection{The Stellar Content of VII~Zw~403}

Fig.~1 contains a wealth of information on the stellar content and
star-formation history of VII~Zw~403, but it needs to be
decoded. Because the spatial distribution of gas and stars, the number
of resolved point sources, and the morphology of the [(V-I)$_o$,
M$_I$$_o$] CMD of VII~Zw~403 are very similar to those of Local Group
Dwarf Irregular Galaxies (dIrr), we let our qualitative interpretation
of the stellar content of VII~Zw~403 be guided by recent results of
ground-based observational and modeling work on the Pegasus Dwarf and
NGC~6822 (Aparicio \& Gallart 1994, Aparicio \& Gallart 1995, Gallart
et al. 1996~a,b,c), including its suggested application to HST data by
Aparicio et al. (1996). We now trace through the evidence for both
recent star formation and an older underlying population.
 
1) That VII~Zw~403 is actively forming stars at the present time is
demonstrated by the strong line emission seen as nebulosity in 
the  H-$\alpha$ image, and also through its
contribution to the V filter in Plate~1. The HII-regions are tracers
of stars with ages $<$10~Myr. 
  
2) Main-sequence (MS) stars and blue supergiants younger than 50~Myr
populate Fig.~1b in a strip extending from [(V-I)$_o$,
M$_I$$_o$]$\approx$[-1, -2] to [0.5, -9].  The young massive stars
also form a distinct band in the [(U-V)$_o$, V$_o$] CMD.
Contrary to what is seen in the [(V-I)$_o$, M$_I$$_o$] CMD of the
Pegasus Dwarf and NGC~6822, the ``red tangle", the clump at
(V-I)$_o$$\approx$1 in Fig.~1b (which includes RGB stars, old and
intermediate-age asymptotic-giant branch (AGB) stars, and
intermediate-age blue-loop (BL) stars) is NOT the single most
prominent feature of the CMD.  Instead the ``blue plume" at
(V-I)$_o$$\approx$0, which contains MS and BL stars, is equally well
populated. So, while VII~Zw~403 clearly exhibits a population of old
stars, it has a much more
abundant population of young stars than these LG-dIrrs. 
We note that in our CMD, especially in the ST filters, the
low-luminosity portion of the ``blue plume" is visibly split into two
distinct color bands.  The bluer band is the MS, and the redder band
might mark the location of the blue turn-around of the blue loops.
The BL contains evolved intermediate-mass and high-mass stars with
ages from 50~Myr to several hundred Myr. A few high-luminosity
(M$_I$$_o$$<$-4) stars are at (V-I)$_o$$\approx$0.3. Such a
population is not apparent in the observed CMDs of the dIrrs, 
possibly removed by foreground-subtraction procedures; however,
very few foreground stars are expected in VII~Zw~403 owing to its high
galactic latitude and the small size of the PC. This population does
appear in the simulated CMDs (e.g. Fig. 5c of Gallart et
al. 1996b). We interpret the objects as high-mass BL stars in
evolutionary stages between the MS and the RSG phase. The [(U-V)$_o$,
V$_o$)] CMD also displays quite a few blue and yellow supergiants.
The single star located at [(V-I)$_o$, M$_I$$_o$]$\approx$[-0.7, -5.5]
is a very bright H$\alpha$ point source within the 
large southern HII region,  
which remains
anomalously blue even after H$\alpha$ subtraction. 
Some blue stars
are within this HII region, but
there are also blue stars with
surrounding HII-regions distributed across the image. 
 
3) The red-supergiant (RSG) strip extends from [(V-I)$_o$,
M$_I$$_o$]$\approx$[1, -4] to [2, -9], and contains stars with ages
similar to those of the stars located in the ``blue plume" of the
CMD. The RSG stars stand out on Plate~1, scattered across the PC
image.
  
4) The AGB stars are located along a band at M$_I$$_o$$\approx$-4.3 to
-5.3 and extending from (V-I)$_o$$\approx$1.5 to 3.0. The age of most of
the AGB stars is several Gyr. A single star located at
(V-I)$_o$$\approx$3.5, M$_I$$_o$$\approx$-6.3 might be a younger (a
few 100~Myr old) AGB star. The mere presence of the ``red tail"
implies that the stellar metallicities for the AGB of VII~Zw~403 are
larger than Z$\approx$0.001, or, for Z$_\odot$=0.02, larger than 1/20
of solar. The red-tail extends almost as far to the red as that of
NGC~6822, which has an oxygen abundance of 1/5 of solar, and for which
the red extension of the AGB fits to the next available grid point
(Z=0.004) in the stellar models of Bertelli et al. (1994).  We suggest
that the upper limit to the stellar metallicities of intermediate-age
stars lies below 1/5 of solar.  It is interesting that the stellar
metallicities of the intermediate-age stars bracket well the oxygen
abundance deduced from the gas.

5) That feature which harbors the truly primeval stellar population of
VII~Zw~403 is the ``red tangle". It extends from [(V-I)$_o$,
M$_I$$_o$]$\approx$[0.5, -2] to [1.5, -4].  The ``red tangle" is
potentially populated by stars with ages spread over the entire
history of the galaxy, from about 0.2 to 15~Gyr.  As discussed by
Aparicio \& Gallart (1994), the ``red tangle" can be split into three
almost vertical bands.  The reddest band is populated by RGB and AGB
stars with ages ranging from 1 to 10~Gyr, the central portion is
populated by the oldest and most metal poor RGB and AGB stars with
ages exceeding 10~Gyr and representing the first generation of stars
formed in this galaxy, while the blue part is populated by BL stars
with ages of a few hundred Myr. Plate~1 illustrates ``Baade's red
sheet" as a background sheet of very numerous, very faint red stars
with a fairly uniform distribution across the face of the galaxy. 

\section{Conclusions}

\noindent$\bullet$ The morphology of the [(V-I)$_o$, M$_I$$_o$] CMD of
VII~Zw~403 resembles that of the well-studied dIrrs NGC~6822 
and the Pegasus Dwarf,
but the ratio of blue-plume to red-tangle stars
is highest for VII~Zw~403, suggesting that its present-day star
formation rate is the highest of the three galaxies. The CMD of
VII~Zw~403 is populated by stars with ages covering a major fraction
of the age of the Universe.
This result emphasizes a point made
previously by Tully et al. (1981), namely that BCDs should not be
considered a separate class of dwarf galaxy, but rather the
high-star-formation extreme in the continuum of dIrr galaxies.

\noindent$\bullet$ The TRGB yields a distance $4.8^{+0.4}_{-0.5}$~Mpc, 
placing
VII~Zw~403 beyond the M~81 group.

\noindent$\bullet$ For the first time, ``Baade's red sheet" is
resolved in a BCD. It is made of AGB and RGB stars, with ages in
excess of several Gyr.  VII~Zw~403 is not a ``young" galaxy forming its
first generation of stars at the present epoch, as is sometimes
proposed for BCDs.

\acknowledgments This work was funded through HST archival 
grant AR-06404.01-95A. UH acknowledges support from the SFB 375.

\clearpage

\figcaption[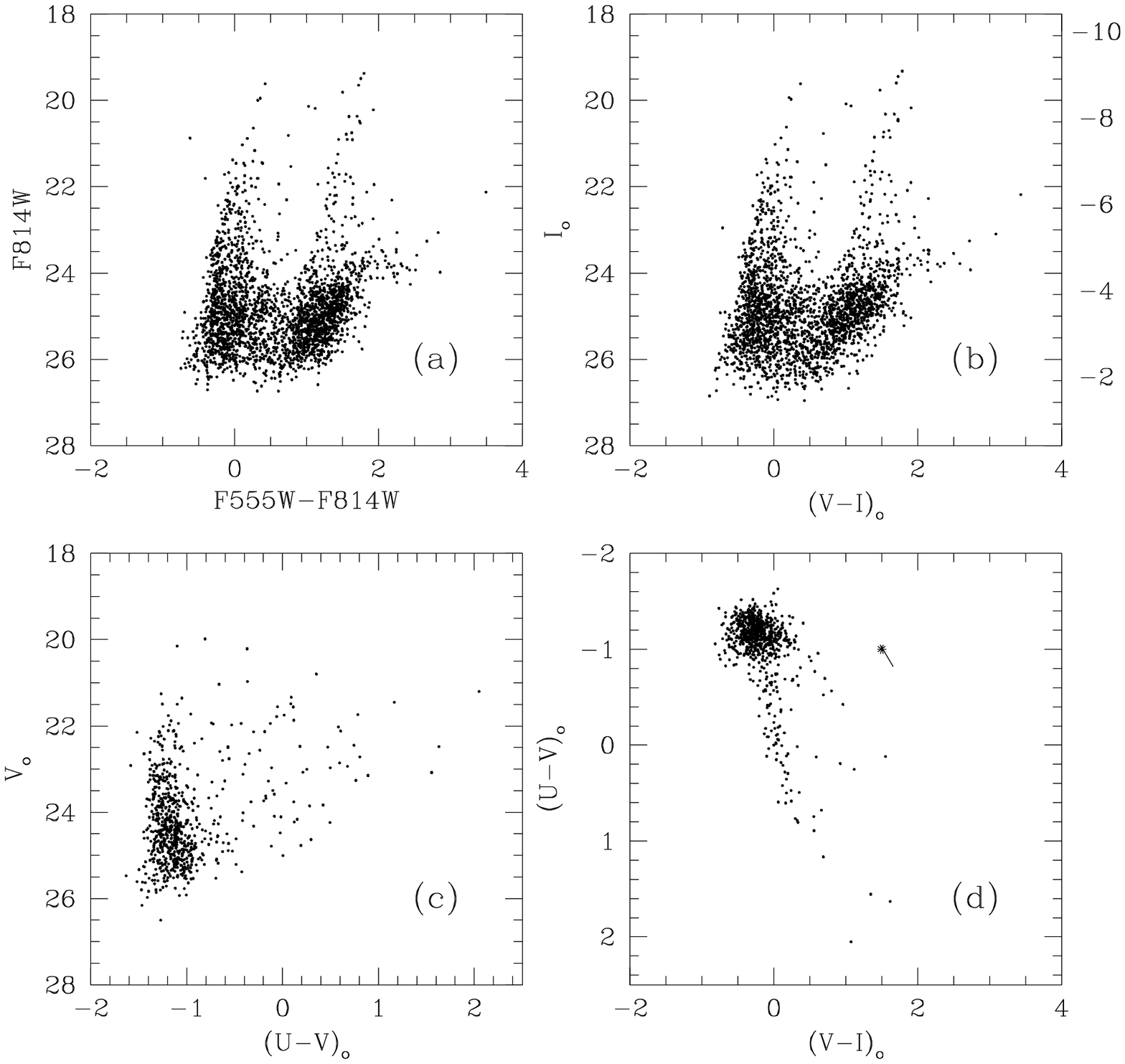]{Color-magnitude and color-color diagrams for
resolved stars in the PC.  ({\it a}) CMD for the
``raw'' photometry for ST filters in the Vega magnitude system. 
({\it b}) The same CMD but with H$_\alpha$ subtraction,
dereddening, and transformation into Johnson-Cousins V and I.  The primary
difference is caused by H$_\alpha$ subtraction -- see text for
discussion.  The absolute M$_I$$_o$ scale on the right is based on
our TRGB distance determination.  
({\it c -- d}) [(U-V)$_o$, V$_o$] CMD and 
color-color diagram.  The clump and linear feature in the color-color
diagram are consistent with an isochrone of a few Myr. We show a
reddening vector for E(B-V)=0.1.}

\figcaption[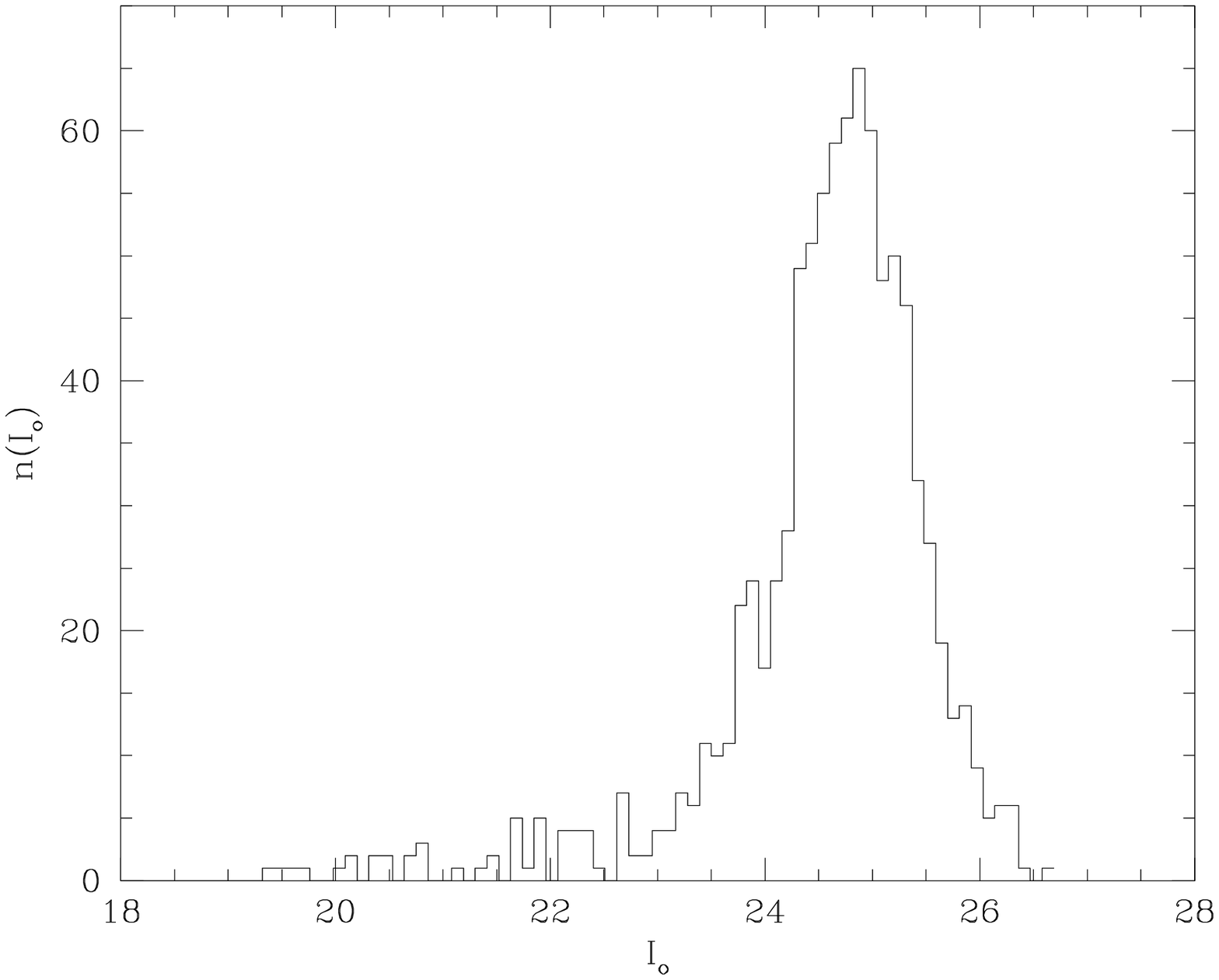]{Luminosity function in I for stars redward of
(V-I)$_o \approx$ 0.7.  The tip of the red giant branch is clearly
visible at I$_o$=24.3.  The smaller step in the distribution at 
I$_o$=23.7 is due to the AGB stars. } 

\notetoeditor{The following is the caption for Plate 1. We couldn't
figure out whether there is a platecaption command, so here it is as
Fig. 3.}

\figcaption[sch_plate1.ps]{Three-color HST PC2 image of the Blue Compact
Dwarf Galaxy VII~Zw~403, showing the
active, central star-forming region of the galaxy.  Images from the
filters F814W, F555W, and F336W were combined using the IRAF COLOR
package. The blue stars clump together, whereas the older, faint red
stars form a uniform background sheet across the image. North is to
the lower right.}

\clearpage

\begin{deluxetable}{lccr}
\tablecaption{WFPC2 Exposures of VII~Zw~403. \label{tbl-1}}
\tablehead{
\colhead{Filter} & \colhead{Exp. time (s)} & \colhead{Data sets}
 & \colhead{Point sources}
 } 
\startdata
F555W & 4200 & u2pq0504t,u2pq0505t,u2pq0506t & 5171 \\
F814W & 4200 & u2pq0507t,u2pq0508t,u2pq0509t & 3478 \\
F336W & 5600 & u2pq050at,u2pq050bt,u2pq050ct,u2pq050dt  &  788 \\
F656N & 2400 & u2pq0501t,u2pq0502t,u2pq0503t &  139 \\
\enddata

\end{deluxetable}

\bigskip

\begin{deluxetable}{lc}
\tablecaption{Errors for TRGB Distance. \label{tbl-2}}
\tablehead{
\colhead{Source} & \colhead{Error (mag)} \cr
}
\startdata 
Residuals from psf fitting \dotfill & $\pm$ 0.05 \cr
Tip measurement    \dotfill         & $\pm$ 0.05 \cr
Galactic extinction \dotfill        & $\pm$ 0.02 \cr
Internal extinction \dotfill        & $\pm$ 0.05 \cr
{\bf Total random errors} \dotfill        & {\bf$\pm$ 0.09} \cr 
\tableline
RR Lyrae distance scale \dotfill     & $\pm$ 0.15 \cr
Absolute magnitude of TRBG \dotfill  & $\pm$ 0.10 \cr
Transformation to I and V \dotfill   & $\pm$ 0.03 \cr
Photometric zero points \dotfill    & $\pm$ 0.02 \cr
{\bf Total systematic errors} \dotfill    & {\bf$\pm$ 0.18} \cr
\enddata

\end{deluxetable}
\end{document}